\documentclass[12pt,english]{article}
\usepackage{amssymb}
\usepackage{mathrsfs, amsmath}
\usepackage[dvips]{graphicx}
\usepackage{setspace}
\usepackage{fancyhdr}
\usepackage{xcolor}
\usepackage{ifpdf}
\usepackage{graphicx}
\usepackage{rotating}
\usepackage{comment,tikz}
\usepackage{url}
\definecolor{darkblue}{rgb}{0,0,0.6}
\usepackage[colorlinks=true,citecolor=darkblue,linkcolor=black,urlcolor=darkblue]{hyperref}
\usepackage{cite}
\usetikzlibrary{patterns}
\usepackage[utf8]{inputenc}

\makeatletter
\newcommand*{\defeq}{\mathrel{\rlap{%
                     \raisebox{0.3ex}{$\m@th\cdot$}}%
                     \raisebox{-0.3ex}{$\m@th\cdot$}}%
                     =}
\makeatother

\DeclareMathOperator{\p}{\partial}
\DeclareMathOperator{\h}{\hspace{-.4mm}\theta}

\newcommand{\be}{\begin{equation}}
\newcommand{\ee}{\end{equation}}

\newcommand{\f}{\frac}

\newcommand{\hz}{\mathcal{H}^{d-1}/\mathbb{Z}}

\newcommand{\bulk}{\text{bulk}}
\newcommand{\ghy}{\text{GHY}}
\newcommand{\ct}{\text{ct}}
\newcommand{\hy}{\mathcal{H}}
\newcommand{\Lp}{L_\phi}
\newcommand{\Lc}{L_\chi}

\begin{document}
\unitlength = 1mm
\ 

\begin{center}

{ \LARGE {\textsc{\begin{center}Detachable circles and temperature-inversion dualities for CFT$_d$ \end{center}}}}

\vspace{0.8cm}
Gary T. Horowitz$^1$ and Edgar Shaghoulian$^{1,2}$

\vspace{.5cm}

$^1$Department of Physics, University of California, Santa Barbara, CA 93106 USA
$^2$Department of Physics, Cornell University, Ithaca, NY 14853 USA

\vspace{1.0cm}

\end{center}

\begin{abstract}
We use a Weyl transformation between $S^1 \times S^{d-1}$ and $S^1 \times \mathcal{H}^{d-1}/\mathbb{Z}$ to relate a conformal field theory at arbitrary temperature on $S^{d-1}$ to itself at the inverse temperature on $\mathcal{H}^{d-1}/\mathbb{Z}$. We use this equivalence to deduce a confining phase transition at finite temperature for large-$N$ gauge theories on hyperbolic space. In the context of gauge/gravity duality, this equivalence provides new examples of smooth bulk solutions which asymptote to conically singular geometries at the AdS boundary. We also discuss implications for the Eguchi-Kawai mechanism and a high-temperature/low-temperature duality on $S^{d-1}$. 

\noindent 
\end{abstract}

\pagebreak
\setcounter{page}{1}
\pagestyle{plain}

\setcounter{tocdepth}{1}
\section{Introduction}
High-temperature/low-temperature dualities have a rich history in physics, beginning with the case of Kramers-Wannier duality in the two-dimensional Ising model. In particular, Kramers-Wannier duality for the square-lattice Ising model maps the theory at a given temperature to itself at some inversely related temperature. Modular invariance of conformal field theories on tori is another famous example of a high-temperature/low-temperature duality which behaves similarly: for  CFT$_2$ \cite{Cardy:1986ie} and CFT$_d$ on special tori 
\cite{Shaghoulian:2015kta, Belin:2016yll} modular invariance acts by mapping the theory at a given temperature to itself at some inversely related temperature.

In this paper we would like to address the question of the existence of high-temperature/low-temperature dualities for CFT$_d$ on curved backgrounds. We will see that this is like asking about Kramers-Wannier duality for non-square lattices, which are not preserved under the duality.
One special case where progress can be made is that of manifolds which can be written as circle fibrations, like odd-dimensional spheres $S^{2n-1}$ presented as a circle fibered over $CP^n$. In this case, there is a proposed modular invariance in the limit where the thermal circle and the fiber circle both become small, with their ratio arbitrary \cite{Shaghoulian:2016gol}.

Motivated by the example of circle fibrations, we wish to consider manifolds with at least one $U(1)$ isometry, for which we can ``detach" the circle corresponding to the $U(1)$ by a Weyl transformation. This is possible whenever there is a hypersurface orthogonal Killing vector with compact orbits. We will focus on the case of finite-temperature conformal field theories on $S^{d-1}$ due to its connection to the local operator content of the CFT. While arbitrary spheres have at least one $U(1)$ isometry, it is not freely acting for even-dimensional spheres. Changing the size of the circle fiber generically gives a manifold with conical singularities. Indeed, the only finite freely acting quotient available for even-dimensional spheres is $\mathbb{Z}_2$. The trick we will use throughout this paper is to detach a circle corresponding to one of these $U(1)$ isometries by a Weyl transformation, after which we are free to treat this direction as Euclidean time and change its periodicity. This allows us to consider a distinct-looking quantization which is fundamentally equivalent, in addition to providing a definition of the conformal field theory on the original orbifold geometry. 

To be concrete, we will use the Weyl transformation
of $S^1\times S^{d-1}$
\be\label{fundequiv}
d\chi^2 + d\h^2 + \sin^2\h d\Omega_{d-3}^2 + \cos^2\h d\phi^2 \longrightarrow \f{d\chi^2+d\h^2 + \sin^2 \h d\Omega_{d-3}^2}{\cos^2 \h}+d\phi^2\,,
\ee
where $\h \in [0,\pi/2]$ for $d>3$ and $\h \in [-\pi/2, \pi/2]$ for $d=3$. Setting $\cos^{-2}\h = r^2 + 1$, the first term on the right becomes $(r^2 +1)d\chi^2 + (r^2 +1)^{-1} dr^2 + r^2d\Omega_{d-3}^2$ which is clearly $\hy^{d-1}/\mathbb Z$. The quotient is due to the $\chi$ coordinate being periodically identified as $\chi \sim \chi + L_\chi$. This space is sometimes called a hyperbolic cylinder. The inverse temperature in the original frame is $L_\chi$, while in the latter frame it is $L_\phi = 2\pi$. 

Physical phenomena are controlled by dimensionless quantities. In the $S^1 \times S^{d-1}$ frame there are two dimensionful scales that can be varied: the length of the $S^1$ and the radius of curvature of the $S^{d-1}$. There is therefore only one dimensionless ratio that can be constructed. By setting the radius of curvature of the $S^{d-1}$ to unity, we denote this ratio as $L_\chi$, which is then just the length of the $S^1$. In the $S^1 \times \hz$ frame there are three dimensionful scales that can be varied: $L_\chi$, $L_\phi$, and the radius of curvature of the hyperboloid (varying $L_\phi$ in the $S^1 \times S^{d-1}$ frame would lead to conical singularities, which will be considered later). This means that there are two independent dimensionless ratios that can be constructed, so the notion of ``high" or ``low" temperature has to be defined more carefully. For our purposes we will signify high or low temperature in terms of the ratio $L_\chi/L_\phi$, with the radius of curvature of the hyperboloid set to unity. This is reasonable since this will often control the physics we are interested in. For example, the confining phase transitions we study will occur at $L_\chi/L_\phi \sim O(1)$. Since our transformation interchanges the role of the thermal circle and the detachable circle, it can be viewed as a high-temperature/low-temperature duality.

One can also use this transformation in reverse. A thermal state of a CFT on $\hy^{d-1}/\mathbb Z$ at any temperature $L_\phi^{-1}$, is equivalent  to the CFT on $S^{d-1}$ with a  new periodicity for the $\phi$ coordinate: $\phi \sim \phi+ 2\pi \alpha$ (with $\alpha = L_\phi/2\pi$). This  gives an orbifold geometry we will call $S^{d-1}_\alpha$, with conical singularities at $\h = 0$ and $\h=\pi/2$ (or $\h=\pm\pi/2$ in $d=3$). For $d=3$ this can be visualized as an American football. 

The temperature-inversion duality implied by the above Weyl transformation is analogous to Kramers-Wannier duality on non-square lattices. For example, the Ising model at a given temperature on a triangular lattice maps under the duality to the Ising model at an inversely related temperature on a honeycomb (hexagonal) lattice. Similarly, our dualities change the spatial background on which the theory is defined. In the context of Kramers-Wannier duality on the triangular lattice, there is a ``star-triangle" relation which allows one to relate the theory on a honeycomb lattice back to the theory on a triangular lattice \cite{Wannier:1945ys}. Combined with Kramers-Wannier duality, this results in a nontrivial high-temperature/low-temperature duality for the theory on a triangular or honeycomb lattice. This duality allows one to determine the critical temperature of the transition on such lattices \cite{Houtappel:1950, Husimi:1950, Wannier:1950zz}. We will comment on a similar attempt to relate our theory on $\hz$ back to the original $S^{d-1}$ in section \ref{startriangle}. 

The equivalence described above make clear two interesting facts. The first is that thermal physics on $S^{d-1}$ can be mapped into thermal physics on $\hz$. In particular, phase transitions in the $S^{d-1}$ frame -- such as the confining transition of large$-N$ gauge theories -- imply phase transitions in the $\hz$ frame. This immediately allows us to exhibit a Hawking-Page phase transition \cite{Hawking:1982dh} for holographic CFTs on $\hz$. To our knowledge this is the first illustration of a Hawking-Page phase transition on hyperbolic surfaces within Einstein gravity. This will be discussed in section \ref{hpsec}. 

Another interesting result of the above equivalence is the ability to write down new smooth geometries that asymptote to conically singular geometries on the AdS boundary. 
(For other examples see, e.g., \cite{Marolf:2010tg,Casini:2011kv,Camps:2016gfs}.)  Weyl transformations on the boundary are given by diffeomorphisms in the bulk, so this allows us to map smooth solutions in the $S^1 \times \hz$ frame into smooth solutions in the $S^1 \times S^{d-1}_\alpha$ frame. In the case where we have $L_\phi \neq 2\pi$, this gives solutions in the $S^1 \times S^{d-1}_\alpha$ frame which are smooth except at the boundary. This will be discussed in section \ref{smoothsec}.

The notion of detachable circles is useful beyond temperature-inversion dualities. As one example, we will discuss implications for the Eguchi-Kawai mechanism in section \ref{qftimp}. For the bulk of this paper we discuss our results in the context of AdS$_4$/CFT$_3$. In section \ref{gensec} we will provide many generalizations: higher dimensions, detaching circles from other manifolds like $S^1 \times \mathcal{H}^{d-1}$ and $S^1 \times \mathbb{R}^{d-1}$, the inclusion of angular momentum, and a high-temperature/low-temperature duality taking place entirely on $S^{d-1}$.
 We conclude with a few open questions in section 7.

\section{The gravitational story}

Since we will often use gauge/gravity duality to explore the phases of holographic CFTs, we start with a discussion of the gravitational solutions we need.  We will use two familiar families of Euclidean solutions of Einstein's equations with a negative cosmological constant\footnote{We set the AdS radius to unity.}:
\be\label{euclschw}
ds^2 = f(r)d\chi^2 + \f{dr^2}{f(r)}+r^2 \left(d \theta^2+\sin^2 \theta d\phi^2\right)\,, \qquad f(r)={r^2} +1-\f{2m}{r}.
\ee
and 
\be\label{euclhyper}
ds^2 = g(\rho)d\phi^2 + \f{d\rho^2}{g(\rho)}+\rho^2\, \f{d\chi^2 + d\tilde \theta^2}{\sin^2 \tilde \h}\,, \qquad g(\rho)={\rho^2} -1-\f{2\mu}{\rho}.
\ee
We require $m\ge 0$ and $\mu \ge \mu_{ext} \equiv - 1/3\sqrt 3$ to avoid curvature singularities. 
Both radial coordinates are restricted to be larger than the zeros of $f$ and $g$: $r\ge r_h$ and $\rho\ge \rho_h$. For $m> 0$ and $\mu > \mu_{ext}$, to avoid conical singularites in the bulk, both $\chi $ and $\phi$ must be periodic in (\ref{euclschw}) with  period $\Lc = 4\pi r_h/(3r_h^2+1)$ for $\chi$ and $L_\phi = 2\pi$ for $\phi$. In (\ref{euclhyper}), only $\phi$ needs to be periodic with period $ \Lp = 4\pi \rho_h/(3\rho_h^2 -1)$.  $\chi$ can be made periodic with arbitrary period $\Lc$  without introducing singularities, and we will consider this general case. In the special case $m=0$, $\Lc$ is arbitrary, and  when $\mu = \mu_{ext}$,  $\Lp$ is arbitrary.
The first metric has manifest spherical symmetry, $SO(3)$, and the second has a hyperbolic symmetry, $SO(2,1)$, which is only broken by the identification of $\chi$. The natural metric\footnote{There is really only a conformal class of boundary metrics, but we will use phrases like  ``natural boundary metric" to refer to the representative one obtains by going to a large radial coordinate in a particular coordinatization of a solution and only rescaling by a power of that radial coordinate.} on the conformal boundaries is $S^1 \times S^2$ for (\ref{euclschw}) and $S^1 \times \hy^2/\mathbb Z$ for (\ref{euclhyper}). 

When $m=0$ and $\mu = 0$, both (\ref{euclschw}) and (\ref{euclhyper}) are globally equivalent and are just  four-dimensional hyperbolic space (or Euclidean AdS). They can be mapped into each other by the coordinate transformation:
\be\label{hypercoord}
 \rho^2 = r^2\sin^2\theta +1, \qquad \cos\tilde \theta = \f{r}{\sqrt{r^2+1}} \cos\theta,
\ee
or its inverse
\be\label{sphercoord}
r^2  = \f{\rho^2}{\sin^2\tilde\h} -1, \qquad \cos^2\theta = \f{\rho^2 \cos^2 \tilde\h}{\rho^2-\sin^2\tilde \h}\,.
\ee
Note that to leading order in large $r$ or $\rho$, $\theta = \tilde\theta$. 

\begin{figure}
\centering
\hspace{-6mm}\includegraphics[scale=0.25]{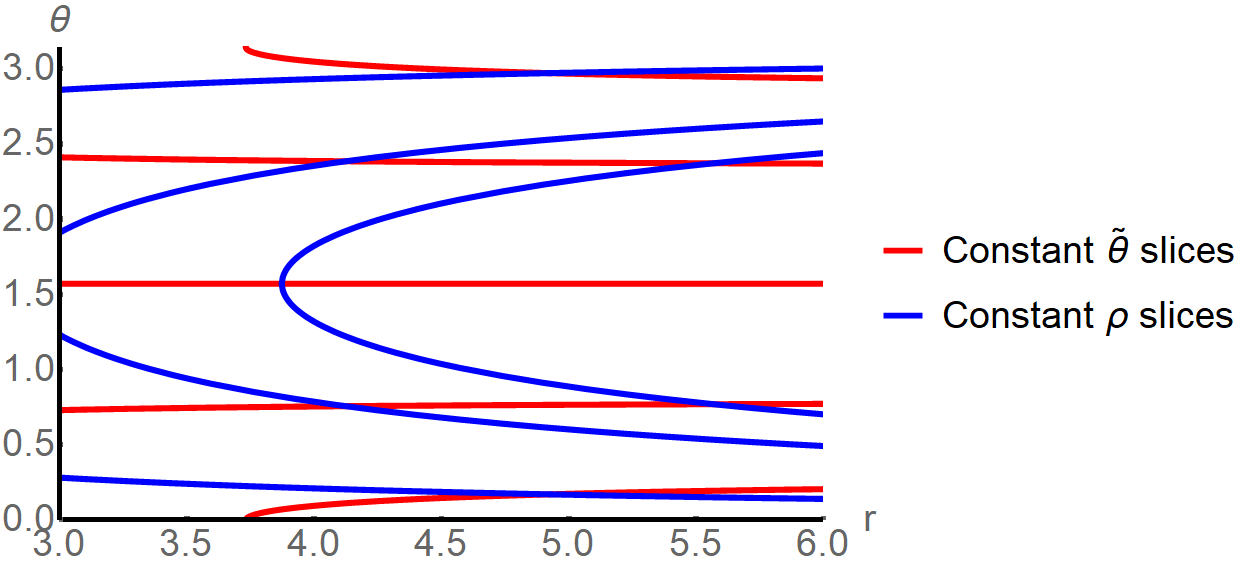}\hspace{4mm}\includegraphics[scale=0.25]{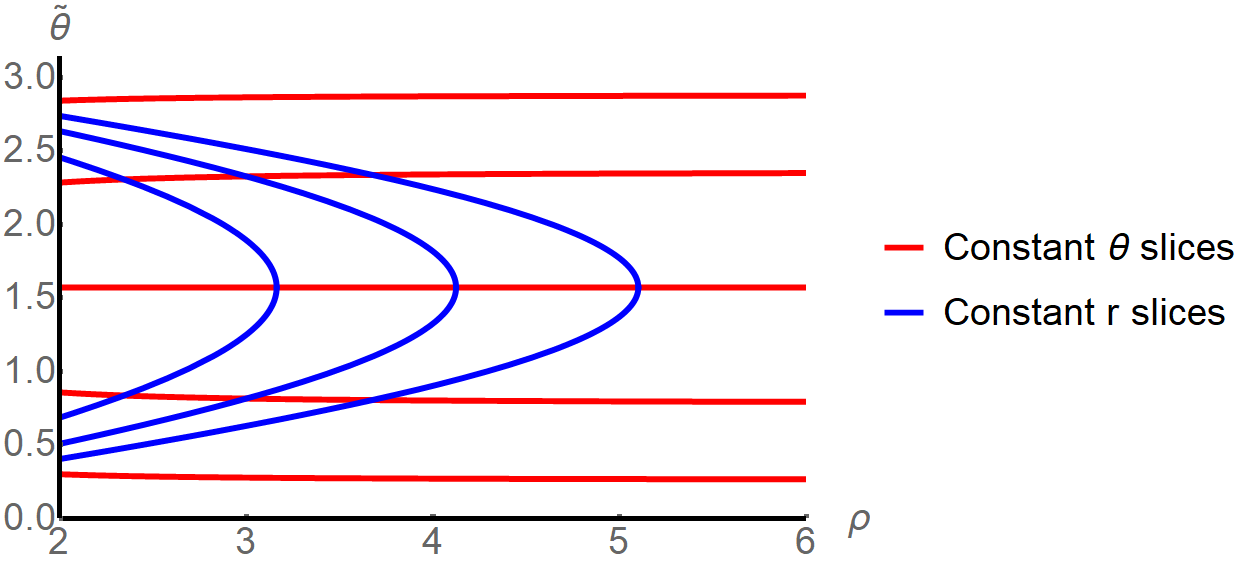}
\caption{ Left: hyperbolic slicing of spherical black hole with $r_h =3$. Right: spherical slicing of hyperbolic black hole with $\rho_h = 2$.} \label{slicings}
\end{figure}

We will use this coordinate transformation even when $m\neq 0$, $\mu \neq 0$. This implements the Weyl transformation (\ref{fundequiv}) (for $d=3$) on the natural boundary metric. The result is that the spherical solutions now acquire a natural asymptotic boundary $S^1 \times \hy^2/\mathbb Z$  and the hyperbolic solutions now acquire a natural asymptotic boundary $S^1 \times S^2$ (possibly with conical singularities). To gain some intuition for these nontrivial slicings, in figure \ref{slicings} we display the hyperbolic slicing of the spherical black hole on the left and the spherical slicing of the hyperbolic black hole on the right. The coordinate ranges in the transformed coordinates are nontrivial as they must be kept outside the black hole horizon. In the first case, the surfaces of constant $\rho$ in the spherical spacetime (\ref{euclschw}) have topology $S^1 \times \mathbb R$ if $\rho > \rho_h$, just like the natural boundary metric. But the circle shrinks to zero size at $\rho = \rho_h$, and for $\rho < \rho_h$ the surface bifurcates into two disconnected surfaces each topologically $\mathbb R^2$. This is illustrated by the $\rho =$ constant line splitting into two in the left-hand diagram in  figure \ref{slicings}. In the second case, the surfaces of constant $r$ are smooth spheres everywhere in the bulk. 

There are two different Lorentzian solutions corresponding to each Euclidean metric depending on which of the two translational symmetries ($\chi$ or $\phi$) we decide to analytically continue. For  solution (\ref{euclschw}),  if we analytically continue $\chi$, we obtain the standard spherical Schwarzschild-AdS black hole. If we analytically continue $\phi$, we obtain an expanding ``bubble of nothing"
\cite{Birmingham:2002st,Balasubramanian:2002am}. The spheres become two dimensional de Sitter spaces written in static coordinates, and the spacetime has a minimal circle.  For solution (\ref{euclhyper}), if we analytically continue $\phi$, we obtain a hyperbolic black hole with horizon $\hy^2/\mathbb Z$. If we analytically continue $\chi$, we obtain a solution with natural boundary metric which is the product of a circle and global $AdS_2$. This is probably the least familiar of the four families of solutions \cite{Birmingham:2002st}. It is the only one which is topologically $\mathbb R^4$.

\subsection{Free energy, total energy, and entropy 
}\label{somecalc}
We now calculate the free energy, total energy, and entropy of the state on $\hy^2/\mathbb Z$ that we obtain by mapping the spherical black hole into hyperbolic slicing using \eqref{hypercoord}.  Since we are in three boundary dimensions, there will be no conformal anomaly, and since our spacetime manifold is compact, there will be no boundary conformal anomaly. The analogous calculations for the state obtained by mapping the  hyperbolic black hole into spherical slicing will be given in section 4.

To begin, we recapitulate the results for the Brown-York stress tensor and free energy for the spherical black holes in their usual slicing. Both objects are regularized by the usual AdS$_4$ holographic counterterms, which with the Gibbons-Hawking-York boundary term gives
\begin{align}
I_E 
&=\underbrace{-\f{1}{16\pi G} \int d^4x \sqrt{g}\left(R+6\right)}_{I_\text{bulk}}+\underbrace{\f{1}{8\pi G} \int_{r=\Lambda} \hspace{-3mm}d^3x \sqrt{\gamma}\, K}_{I_\text{GHY}} + \underbrace{\f{1}{8\pi  G} \int_{r=\Lambda} \hspace{-3mm}d^3x \sqrt{\gamma}\left(2+\f{1}{2}R[\gamma]\right)}_{I_\text{ct}}.
\end{align}
Vacuum solutions satisfy $R = -12$, giving a bulk contribution 
\be
I_{\textrm{bulk}}=\f{3}{8\pi G}\int d^4 x \sqrt{g}\,
\ee
in all cases.

The spherical Schwarzschild-AdS black hole, in its original slicing with coordinates $(r,\theta)$, has an on-shell action that is evaluated by picking a cutoff $r=\Lambda$, summing all three contributions, and taking $\Lambda \rightarrow \infty$ at the end:
\begin{align}
I_{\textrm{bulk}}&= \f{L_\chi}{2 G}\left( \Lambda^3-r_h^3 +\cdots \right),\\
I_{\text{ct}}&=\f{L_\chi}{2 G}\left(2  \Lambda^3+2  \Lambda -2 m  +\cdots \right),\\
I_{\text{GHY}}&=\f{L_\chi}{2 G}\left(-3  \Lambda^3-2 \Lambda +3m +\cdots\right).\\
\implies I_E &= \lim_{\Lambda\rightarrow \infty} \left(I_{\text{bulk}}+I_{\text{ct}}+I_{\text{GHY}} \right)=\frac{L_\chi  \left( m-r_h^3\right)}{2  G } = \frac{L_\chi  \left( r_h-r_h^3\right)}{4  G }\,. \label{finalsphere}
\end{align}
where we have used $m = r_h(1+r_h^2)/2$ in the last step. This is the standard answer for the free energy of the Schwarzschild-AdS black hole in four dimensions. 

We now want to reproduce this answer in hyperbolic coordinates by applying the diffeomorphism \eqref{hypercoord} to the bulk solution. This is not trivially the same calculation since we will now pick a cutoff $\rho = \widetilde{\Lambda}$ instead of the $r=\Lambda$ cutoff just chosen. The coordinate ranges in $\rho, \tilde{\theta}$ have to be chosen so that they do not cover the region behind the horizon $r_h$ (see the left-hand-side of figure \ref{slicings}). We also want to put cutoffs at $\tilde{\theta} = \epsilon$ and $\tilde{\theta} = \pi - \epsilon$ to cut off the divergence from the infinite hyperbolic space. Altogether this gives
\begin{align}
I_{\textrm{bulk}}&= \f{L_\chi \cot \epsilon}{2 G}\left(\widetilde{\Lambda}^3-1-\f{r_h^3}{\cot \epsilon} +\cdots \right),\\
I_{\text{ct}}&=\f{L_\chi\cot \epsilon}{2 G}\left(2  \widetilde{\Lambda}^3-2  \widetilde{\Lambda} -2 m \sin \epsilon +\cdots \right),\\
I_{\text{GHY}}&=\f{L_\chi\cot \epsilon}{2 G}\left(-3  \widetilde{\Lambda}^3+2 \widetilde{\Lambda} +3m \sin \epsilon+\cdots\right).\\
\implies I_E &= \lim_{\widetilde{\Lambda}\rightarrow \infty} \left(I_{\text{bulk}}+I_{\text{ct}}+I_{\text{GHY}} \right)=\frac{L_\chi  \left( m \cos \epsilon -r_h^3 - \cot \epsilon\right)}{2  G }\,.
\end{align}
We see that there is a divergent piece as $\epsilon \rightarrow 0$. This is independent of the mass $m$ and comes simply from the infinite hyperbolic space.  The reason for this residual divergence is that even for finite $\rho$, the hyperbolic slices reach the boundary at infinity, so a cut-off at $\rho = \widetilde{\Lambda}$ does not remove all infinities in the action. Since we want the action for pure AdS to vanish, we subtract the $\cot\epsilon$ term.  With this prescription we can take $\epsilon \rightarrow 0$ and recover \eqref{finalsphere}. 

We now compute the total energy by integrating the Brown-York stress tensor. In the original spherical slicing, the metric on a large $r$ surface is 
\be
ds^2 = r^2\left(d\chi^2+d\h^2 + \sin^2\h d\phi^2\right).
\ee
The Euclidean stress tensor  is
\begin{gather}
 T_\mu^\nu = \text{diag}\left(T_{\chi}^{\chi}, T_{\h}^{\h}, T_{\phi}^{\phi}\right) = \text{diag}\left(\f{- m}{4\pi  G r^3},\, \f{ m}{8\pi G r^3},\, \f{ m}{8 \pi G r^3}\right),\\
T_{\mu\nu} = \text{diag}\left(\f{-m}{4\pi  G  r},\, \f{ m}{8\pi G r},\, \f{ m \sin^2\h}{8 \pi G r}\right).
\end{gather}
In hyperbolic coordinates the metric is rescaled by $\sin^{-2}\theta$, so by conformal invariance the stress tensor is rescaled as well:
\begin{gather}
 T_\mu^\nu[S^1 \times \hy^2/\mathbb{Z}] = \sin^3 \h T_\mu^\nu[S^1 \times S^2] = \sin^3 \h \text{diag}\left(\f{- m}{4\pi  G r^3},\, \f{ m}{8\pi G r^3},\, \f{ m}{8 \pi G r^3}\right),\\
T_{\mu\nu}[S^1 \times \hy^2/\mathbb{Z}] = \text{diag}\left( \f{-m \sin\h}{4\pi  G r},\, \f{ m \sin \h}{8\pi G r},\,\f{ m \sin^3 \h}{8\pi G r}\right)\,.
\end{gather}
This result can also be obtained by an explicit calculation of the Brown-York stress tensor in hyperbolic slicing. Sticking a minus sign into $T_{\phi \phi}$ lets us interpret $\phi$ as time and gives an energy density
\be\label{totalE}
T_{\phi \phi}[S^1  \times \hy^2/\mathbb{Z}] = \f{- m \sin^3\h}{8\pi G r}\implies E[S^1 \times \hy^2/\mathbb{Z}] = \int T_{\phi \phi}[S^1 \times \hy^2/\mathbb{Z}] \f{r \,d\h d\chi}{\sin^2\h } = \f{- L_\chi m}{4\pi G}\,.
\ee
Note that this energy is always negative. This  is similar to the AdS soliton
 \cite{Horowitz:1998ha}.

We next use the usual Bekenstein-Hawking formula $S=A/(4G)$ to compute the entropy.
Since Euclidean time is now $\phi$, the  thermal circle caps off at $\theta = 0$ and $\theta = \pi$ throughout the entire spacetime. This is just the de Sitter entropy on each radial slice coming from quantizing along the azimuthal angle of the sphere. This horizon is just like the extremal surfaces used in the Ryu-Takayanagi formula to calculate entanglement entropy on the boundary. Its area is  infinite, but it can be regulated using the usual proceedure and has a finite piece given by 
\be\label{entropy}
S=-\f{L_\chi r_h}{2G}<0\,.
\ee

In the hyperbolic slicing, the inverse temperature is $L_\phi = 2\pi$. We can now combine this with $E$ and $S$ to give an independent calculation of the free energy:  $-\log Z = I_E = 2\pi E - S$.
Combining \eqref{totalE} and \eqref{entropy},  we have
\be
-\log Z[S^1  \times \hy^2/\mathbb{Z}]=2\pi F =2\pi E-S=   \frac{L_\chi  \left( r_h-r_h^3\right)}{4  G }\,,\ee
where we have again used $m = r_h(1+r_h^2)/2$. This agrees precisely with the previous method of calculation. 

These values of the energy and entropy can also be obtained from $-\log Z = I_E$ by the thermodynamic formulas $S= (1-L_\phi \partial_{L_\phi})\log Z$ and $E = -\partial_{L_\phi} \log Z$. Since we are differentiating with respect to $L_\phi$ this will introduce conical singularities into the spacetime and must therefore be treated with care. The appropriate prescription is to consider the solution with arbitrary $L_\phi$, which for $L_\phi \neq 2\pi$ requires a delta-function source to support. This delta function feeds into the on-shell action through the Ricci scalar. Altogether, after subtracting some pieces linearly divergent in $\Lambda$, the same results for $E$ and $S$ can be obtained.

\section{A thermal phase transition on $S^1\times\mathcal{H}^{d-1}/\mathbb{Z}$}
\label{hpsec}

Before considering the gravitational story
which is appropriate at strong coupling, let us quickly remark on free large-$N$ conformal gauge theories. These are expected to undergo a confinement/deconfinement phase transition at finite temperature on $S^{d-1}$, with calculable critical temperatures available for various theories including $N=4$ super Yang-Mills and ABJM theory \cite{Sundborg:1999ue, Aharony:2003sx, Nishioka:2008gz}. In the free limit this temperature coincides with the Hagedorn temperature. These results can now immediately be mapped to a confinement/deconfinement phase transition at finite temperature on $\hz$ due to our Weyl transformation (which will be generalized to arbitrary dimension in section \ref{higherdim}). 

  We have seen that for a CFT on $S^1 \times \hy^2/\mathbb Z$ there are
  four possible bulk solutions. In addition to the usual extremal and nonextremal hyperbolic solutions (\ref{euclhyper}), one can take the spherical solution (\ref{euclschw}) with either $m=0$ or $m>0$ and write it in the coordinates (\ref{hypercoord}). In this section we will discuss which one dominates the canonical ensemble and deduce a phase transition.

Since the solution  (\ref{euclschw})  is the basis for the  Hawking-Page phase transition \cite{Hawking:1982dh}, we first review this transition.  Recall that for large $\Lc$, the only solution of the form (\ref{euclschw})  has $m=0$.  For small $\Lc$, there are two solutions with $m>0$ as well as a solution with $m=0$.  It is clear from \eqref{finalsphere} that the Euclidean action becomes negative for $r_h > 1$ corresponding to $\Lc < \pi$.  So  for $\Lc > \pi$, the $m=0$ solution has lowest action while for $\Lc < \pi$  the solution with $r_h > 1$  has lowest action. The standard gravitational interpretation, viewing $\chi$ as Euclidean time, is that at high temperature, the  canonical ensemble is dominated by a large black hole while for low temperature, it is dominated by a gas of particles in AdS. In terms of the dual field theory on $S^1 \times S^2$, one has a deconfined phase for  $\Lc < \pi$ and a confined phase for $\Lc > \pi$ \cite{Witten:1998zw}.

We can now reinterpret this as a phase transition for CFTs on $S^1\times\mathcal{H}^{2}/\mathbb{Z}$.   Since $\Lp = 2\pi$,
there are only three different bulk solutions since the  $\mu =0 $ hyperbolic black hole is equivalent to the the $m=0$ solution.  At nonzero temperature, the $\mu=\mu_{ext}$ solution always has higher free energy than the $\mu > \mu_{ext}$ one. (We will see this explicitly in the next section.) So we are left with the $m=0$ and $m>0$ spherical solutions, which are just the ones used for the Hawking-Page transition.  Since the roles of $\chi$ and $\phi$ are now reversed we are led to the following  conclusion:  {\it a CFT on $S^1\times\mathcal{H}^{2}/\mathbb{Z}$ at fixed temperature $T = 1/2\pi$ undergoes a phase transition when the identification of the hyperboloid is varied}.  Note that $\Lc$ is the minimum size of the noncontractible circle in $\mathcal{H}^{2}/\mathbb{Z}$. When $\Lc > \pi$, this circle never pinches off in the bulk. It is analogous to the deconfined phase. The bulk is locally $AdS_4$ and is just the $\mu=0$ hyperbolic black hole (with an identification on the hyperboloid). When $\Lc < \pi$, the circle does pinch off in the bulk, and the theory is in a phase analogous to the confined phase. Note that the interchange of the thermal circle with a spatial circle also interchanges the confining and deconfining phases.

Since the confined phase is described by the spherical solution (\ref{euclschw}), it breaks the hyperbolic symmetry. We thus have an exact analytic description of a localized excitation on the hyperboloid. Since we are now thinking of $\phi$ as Euclidean time, the bulk solution should be thought of as a bubble of nothing. 

The existence of a phase transition at a fixed temperature is not surprising. A familiar example is the confining-deconfining transition for a CFT on $T^2 \times R$. Viewing one $S^1$ as Euclidean time with period $\beta$, and denoting the length of the other circle as $L$, then for fixed $\beta$ there is a phase transition as we change $L$: for $L>\beta$, the CFT is deconfined and the bulk dual is the planar black hole, while for $L< \beta$,  the system is confined and the bulk dual is the AdS soliton. Of course, this phase transition extends to other temperatures, and we expect the same will be true for the phase transition on $S^1\times\mathcal{H}^{2}/\mathbb{Z}$ away from $T = 1/2\pi$.

\begin{figure}
\centering
\includegraphics[scale=0.42]{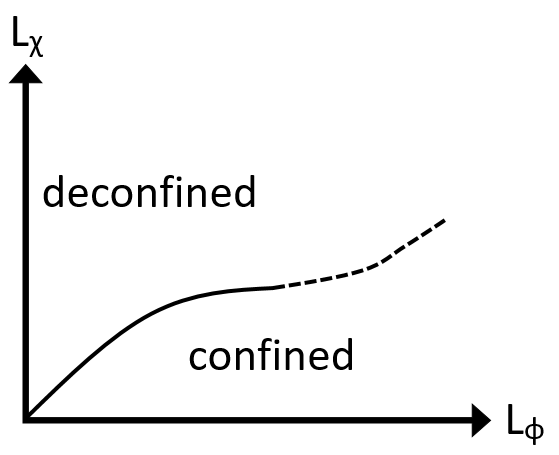}
\caption{A conjectured phase diagram on $S^1 \times \hz$.  
The solid curve meets the dashed curve at the calculable transition point $L_\phi = 2\pi$, $L_\chi = {2\pi}/(d-1)$.} \label{phasediag}
\end{figure}

A conjectured phase diagram is shown in figure \ref{phasediag}. If we start in the deconfined phase and increase the temperature, we expect to remain in the deconfined phase. The natural bulk dual of this is the hyperbolic black hole (with an identification on the hyperboloids). If we start in the confined phase and increase the temperature, we expect to eventually enter the deconfined phase which is again described by the hyperbolic black hole. Since we do not know a nonsingular bulk solution describing the confined phase away from $T = 1/2\pi$, we cannot calculate the precise temperature of this transition. The required solution would have boundary $S^1\times\mathcal{H}^{2}/\mathbb{Z}$ where the $S^1$ has length different than $2\pi$ and the circle in the $\mathcal H^2$ is  contractible in the bulk.  Going the other way, if we lower the temperature starting in the confined phase, we expect to remain in the confined phase. 
At small $L_\chi$ and small $L_\phi$, we expect the curvature of the hyperboloid to become irrelevant and the transition to look like that of a flat cylinder, i.e. to occur at $L_\chi = L_\phi$. These facts are captured by the line drawn at a forty-five degree angle near the origin in figure \ref{phasediag}. We have drawn the simplest curve that interpolates between this line at the origin and the calculable transition point at $L_\phi = 2\pi$, $L_\chi = \pi$. If we increase $L_\chi$ while in the confined phase, we expect to deconfine since we will reach the unquotiented hyperbolic black hole as $L_\chi \rightarrow \infty$, for which we are always deconfined. To reflect this we have continued the curve 
(with dashes) in an arbitrary way which still allows for this behavior. In particular the dashed curve cannot asymptote to a vertical line (although it can asymptote to a horizontal line). We will see in section \ref{higherdim} that this phase transition exists on $S^1 \times \hz$ for all $d\ge 3$,  with a calculable transition point at $L_\phi = 2\pi$, $L_\chi = {2\pi}/(d-1)$ as one varies $L_\chi$. Thus the phase diagram again looks like figure 2.
 
A similar phase transition for CFTs on $dS_2 \times S^1$ was discussed in \cite{Marolf:2010tg} (and on $dS_3 \times S^1$ in \cite{Hutasoit:2009xy}) using the Hawking-Page transition and viewing $\phi$ as Euclidean time. In that context, the temperature $T = 1/2\pi$ is fixed as the natural de Sitter temperature (since we have set $H=1$). The authors of \cite{Marolf:2010tg} tried unsuccessfully to  find smooth bulk solutions corresponding to $dS_2 \times S^1$ at different temperatures to see how the phase transition extends away from this temperature.  We will see how to use our Weyl transformation to find one such solution in the next section.

\section{Constructing smooth bulk solutions with conical boundaries}\label{smoothsec}

Suppose we start with the hyperbolic solution (\ref{euclhyper}) and introduce the spherical coordinates (\ref{sphercoord}).  The metric on constant $r$ surfaces, as $r\to \infty$, is now $d\chi^2 + d  \tilde \theta^2 + \sin^2 \tilde \h d\phi^2$, but since the period of $\phi$  is not in general $2\pi$, the sphere has conical singularities at the poles and (for $\mu >0$) resembles an American football. We will denote this by $S^2_\alpha$, where the period of $\phi$ is $2\pi \alpha$. Since the bulk is perfectly nonsingular, this provides a nonsingular gravitational dual for a CFT$_3$  on a space with conical singularities: $S^1\times S^2_\alpha$.
We actually have two nonsingular bulk solutions for each boundary $S^1\times S^2_\alpha$: one with $\mu > \mu_{ext}$ and one with $\mu = \mu_{ext}$. 
As we noted in section 2,  for $\mu > \mu_{ext}$, the surfaces of constant $r$ in the geometry (\ref{euclhyper}) are smooth spheres everywhere in the bulk. It is only in the limit $r\to \infty$ that they become singular.  

There are previous examples of nonsingular bulk metrics having boundary metrics with conical singularities. In AdS$_3$/CFT$_2$ this is easy since a conical singularity can be introduced in any two dimensional metric by a Weyl transformation.  In  \cite{Marolf:2010tg}, the metric (\ref{euclhyper}) was used (without identification of the $\chi$ coordinate) to provide a nonsingular bulk  for a boundary $S^3$ with conical singularities. This is possible since $S^3$ is Weyl-equivalent to $S^1\times \hy^2$. This was interpreted as describing a state of a CFT on dS$_3$ with temperature different from the de Sitter temperature. 

The nonsingular bulk descriptions of  $S^1\times S^2_\alpha$ appear to be new and have two interesting consequences. First, viewing the $S^1$ as a thermal circle, they provide a  description of a CFT at any temperature on $S^2_\alpha$. Since we have two nonsingular bulk solutions for each boundary $S^1\times S^2_\alpha$, it is natural to ask which one
dominates the canonical ensemble. Since the  $\mu > \mu_{ext}$ solution has lower free energy than the $\mu = \mu_{ext}$ solution, it will always dominate. In particular, in the limit of zero temperature, the ground state corresponds to the $\mu > \mu_{ext}$ solution. The fact that $\mu = \mu_{ext}$ has the lowest energy among hyperbolic black holes is not relevant here since we are measuring energy with respect to a different Killing field. As we will see by calculating this energy in  \eqref{hypenergy} the $\mu> \mu_{ext}$ solution always has lower energy since $E \propto -\mu$ (but of course all energies are infinite). 

As a second consequence, viewing $\phi$ as Euclidean time, the boundary describes a static patch of $dS_2 \times S^1$ with a temperature different than the standard de Sitter temperature, $T=1/2\pi$. This is the solution sought by the authors of  \cite{Marolf:2010tg}. However, it is not enough to complete their phase diagram since the $S^1$ is never contractible in the bulk. (It corresponds to the $\chi$ circle in the original hyperbolic black hole.)

\subsection{Free energy of hyperbolic black hole in spherical slices}
The hyperbolic Schwarzschild-AdS black hole, in its original slicing with coordinates $(\rho,\tilde{\theta})$, has an on-shell action that is evaluated by picking a cutoff $\rho=\widetilde{\Lambda}$, summing all three contributions, and taking $\widetilde{\Lambda} \rightarrow \infty$ at the end:
\begin{align}
I_\bulk&=\f{L_\phi\text{Vol}\left[\mathbb{H}^2/\mathbb{Z}\right]}{8\pi G}\left(\widetilde{\Lambda}^3-\rho_h^3+\cdots\right),\\
I_\ct&=\f{L_\phi\text{Vol}\left[\mathbb{H}^2/\mathbb{Z}\right]}{8\pi G}\left(2\widetilde{\Lambda}^3-2\widetilde{\Lambda}-2\mu+\cdots\right)\,,\\
I_\ghy&=\f{L_\phi\text{Vol}\left[\mathbb{H}^2/\mathbb{Z}\right]}{8\pi G}\left(-3\widetilde{\Lambda}^3+2\widetilde{\Lambda}+3\mu+\cdots\right)\,,\\
\implies-\log Z&= \lim_{\widetilde{\Lambda}\rightarrow \infty}\left(I_\bulk+I_\ct+I_\ghy\right) = \f{L_\phi  \text{Vol}\left[\mathbb{H}^2/\mathbb{Z}\right]}{8\pi G}\left(\mu-\rho_h^3\right)\\
&\hspace{49mm} = \f{L_\phi  \text{Vol}\left[\mathbb{H}^2/\mathbb{Z}\right]}{16\pi G}\left(-\rho_h-\rho_h^3\right) \,,\label{finalhyper}
\end{align}
where we have used $\mu = \rho_h (-1+\rho_h^2)/2$ in the last step. This is the standard answer for the free energy of the hyperbolic Schwarzschild-AdS black hole in four dimensions. Notice that all CFT UV divergences (terms scaling as positive powers of $\widetilde\Lambda$) cancel due to the regularization, but there remains a CFT IR divergence from the infinite volume of $\mathcal{H}^2/\mathbb{Z}$.

We now want to reproduce this answer in spherical coordinates by applying the diffeomorphism \eqref{hypercoord} to the bulk solution. This is not trivially the same calculation since we will now pick a cutoff $r = \Lambda$ instead of the $\rho=\widetilde{\Lambda}$ cutoff just chosen. The coordinate ranges in $r,\theta$ have to be chosen so that they do not cover the region behind the horizon $\rho_h$ (see the right-hand-side of figure \ref{slicings}). Notice also that our Weyl transformation maps the IR into the UV, so we should expect the untamed IR divergence of the hyperbolic black hole to appear as an untamed UV divergence in spherical slicing. To isolate this, we will not perform the $\theta$ integrals until the end, since this is where the new divergence should be hiding. Altogether we have 
\begin{align}
I_{\textrm{bulk}}&= \int d\theta \,\f{L_\phi L_\chi \sin \theta}{8\pi  G}\left(\Lambda^3-\f{\left(\rho_h^2-1\right)^{3/2}}{\sin^3\theta} +\cdots \right),\\
I_{\text{ct}}&=\int d\theta \,\f{L_\phi L_\chi \sin \theta}{8\pi  G}\left(2  \Lambda^3+2  \Lambda -\f{2\mu}{ \sin^3 \theta} +\cdots \right),\\
I_{\text{GHY}}&=\int d\theta \,\f{L_\phi L_\chi \sin \theta}{8\pi  G}\left(-3  \Lambda^3-2\Lambda +\f{3\mu}{ \sin^3 \theta}+\cdots\right).\\
\implies I_E &= \lim_{\Lambda\rightarrow \infty} \left(I_{\text{bulk}}+I_{\text{ct}}+I_{\text{GHY}} \right)=\int d\theta \,\f{L_\phi L_\chi \left(\mu-\left(\rho_h^2-1\right)^{3/2}\right)}{8\pi  G \sin^2\theta}\,.
\end{align}
Notice that the combination $L_\chi/\sin^2\theta$ is precisely the appropriate determinant which gives Vol$[\mathcal{H}^2/\mathbb{Z}]$ when integrated over $\theta \in [0,\pi]$. In this case, we have cutoffs near $\theta=0$ and $\theta = \pi$ set by $\rho_h$ and $\Lambda$, so integrating the final answer and using $\mu = \rho_h (-1+\rho_h^2)/2$ gives 
\be
I_E = \f{L_\phi L_\chi \Lambda \left(2-2\rho_h^2+\rho_h\sqrt{\rho_h^2-1}\right)}{8\pi  G}\,.
\ee
As discussed earlier, the IR divergence Vol$[\mathcal{H}^2/\mathbb{Z}]$ in hyperbolic coordinates has mapped into a UV divergence in spherical coordinates. Notice that this transformation tames the infrared divergence of the topological ($\mu=0$) black hole. 

The energy can be computed using the Brown-York stress tensor. For the hyperbolic black hole with boundary $S^1 \times \mathcal{H}^2/\mathbb{Z}$ we have
\begin{gather}
ds^2 = \f{d\chi^2+d\h^2}{\sin^2\h} + d\phi^2\\
\implies T_\mu^\nu = \text{diag}\left(T_{\chi}^{\chi}, T_{\h}^{\h}, T_{\phi}^{\phi}\right) = \text{diag}\left(\f{ \mu}{8\pi  G r^3},\, \f{\mu}{8\pi G r^3},\, \f{- \mu}{4 \pi G r^3}\right),\\
T_{\mu\nu} = \text{diag}\left(\f{\mu}{8\pi  G r \sin^2 \h},\, \f{\mu}{8\pi G r \sin^2 \h},\, \f{- \mu }{4 \pi G  r}\right)\label{hypstress}
\end{gather}
which maps to the $S^1 \times S^2_\alpha$ frame as 

\begin{gather}
ds^2 = {d\chi^2+d\h^2}+ \sin^{2}\h d\phi^2\\
\implies T_\mu^\nu[S^1 \times S^2_\alpha] = \sin^{-3} \h T_\mu^\nu[S^1 \times \hy^2/\mathbb{Z}] = \sin^{-3} \h \text{diag}\left(\f{\mu}{8\pi  G r^3},\, \f{\mu}{8\pi G r^3},\, \f{-\mu}{4 \pi G r^3}\right),\\
T_{\mu\nu}[S^1 \times S^2_\alpha] = \text{diag}\left( \f{\mu }{8\pi  G r \sin^3 \h},\, \f{\mu }{8\pi G r \sin^3 \h},\,\f{- \mu }{4\pi G  r \sin \h},\right)
\end{gather}
which is validated by an explicit calculation in $S^1 \times S^2_\alpha$ slicing. Sticking a minus sign into $T_{\chi\chi}$ to interpret $\chi$ as time gives an energy 
\be
E[S^1 \times S^2_\alpha] = \int T_{\chi\chi}[S^1 \times S^2_\alpha] r\sin\h d\h d\phi = -\f{\mu L_\phi}{4\pi G} \int_0^\pi \f{d\h}{\sin^2\h}\,.\label{hypenergy}
\ee
This infinity is reasonable; a constant energy density on the hyperboloid (and therefore constant pressure) maps to an energy density on the sphere that diverges at the north and south poles $\h=0,\pi$. This infinity has no finite piece in it, so it is reasonable to regulate the energy to vanish. Then, since the entropy also vanishes, we are left with $\log Z  = 0$, agreeing with the previous results showing no finite piece in the free energy.

\section{Implications for the Eguchi-Kawai mechanism on curved manifolds}\label{qftimp}
The trick of performing a Weyl transformation to isolate a circle factor within the Euclidean manifold of the theory is useful beyond modular relations. As another concrete application, we will discuss here a simple implication for the Eguchi-Kawai mechanism \cite{Eguchi:1982nm}, also known as large-$N$ volume independence, for conformal large-$N$ gauge theories. 

While there are many variants and generalizations, the simplest manifestation of the Eguchi-Kawai mechanism applies to large-$N$ quantum gauge theories with a one-form center symmetry. When defined on a Euclidean manifold with $S^1$ factors, and when translation invariance and center symmetry along a given $S^1$ factor is preserved, then appropriate observables (like the free energy density and certain correlation functions) become independent of the size of the $S^1$ at leading order in $N$. For a longer summary see \cite{Kovtun:2007py}. 

The argument in this section will proceed as follows. We will first use the standard Eguchi-Kawai mechanism  to learn about our CFT in the frame with a detached circle, after which we will use our Weyl transformation to deduce a generalization of the Eguchi-Kawai mechanism to curved manifolds. More sharply, in conformal large-$N$ gauge theories with a center symmetry -- like holographic CFTs -- we can use our Weyl transformation to isolate circle factors and deduce independence of appropriate observables on the size of the $S^1$ (as long as center symmetry and translation invariance are not spontaneously broken). We can for example consider our duality between $S^1 \times S^{d-1}_\alpha$ and $S^1_\alpha \times \hz$. Varying the circle in the latter Weyl frame corresponds to a variation of the azimuthal angle in $S^{d-1}_\alpha$ in the original frame. The standard Eguchi-Kawai mechanism in the latter frame thus implies a curved-space generalization of the Eguchi-Kawai mechanism in the former frame. The curved space is an orbifold geometry, so leads to an orbifold generalization of the notion of ``topological large-$N$ volume independence" introduced in \cite{Shaghoulian:2016xbx}.\footnote{Topological large-$N$ volume independence is a proposed generalization of the Eguchi-Kawai mechanism to curved manifolds. It says that independence of the size of factor $S^1$'s can be generalized to independence of size of curved manifolds under smooth, topological changes in volume, like lensing of odd-dimensional spheres.}

At present this seems a bit formal, so let us use the Eguchi-Kawai mechanism to learn about the holographic phase structure on $S^1 \times S^{2}_\alpha$. We have already argued that the phase on
\be
ds^2 = d\phi^2 + \f{d\chi^2 + d\h^2}{\sin^2\h}\,,\qquad \phi \sim \phi + L_\phi, \qquad \chi \sim \chi + L_\chi
\ee
should be confined for $L_\phi = 2\pi$ and $L_\chi < \pi$, simply by mapping the phase structure and thermal Wilson loop order parameter from the $S^1 \times S^{2}_\alpha$ frame. Physically, decreasing the temperature while in the confined phase should keep you in the confined phase. This means that for fixed $L_\chi  < \pi$, increasing $L_\phi$ past $2\pi$ preserves center symmetry along the $\phi$ cycle. But as discussed before, we no longer have a smooth bulk saddle in this region of parameter space from which to read off a free energy density. This is because the smooth confined phase saddle at $L_\phi= 2\pi$ is the spherical Schwarzschild-AdS black hole, which develops an orbifold singularity for $L_\phi \neq 2\pi$. Nevertheless, by the standard Eguchi-Kawai mechanism, we know that the free energy density should be independent of the periodicity of $L_\phi$ as it is increased. This means that we can deduce the free energy density even without a smooth bulk saddle! Furthermore, the free energy density is given accurately by the singular Schwarzschild-AdS geometry with the rule that there is no additional contribution from the orbifold singularities on the boundary geometry (the usual delta-function stress-energy source term and response in the Ricci scalar are to be included and cancel against each other in a bulk on-shell calculation). 

By Weyl equivalence this tells us about the phase structure on $S^1 \times S^2_\alpha$. In particular, we have topological volume independence with respect to the family of orbifold geometries $S^2_\alpha$.  This is the orbifold generalization of the Eguchi-Kawai mechanism alluded to earlier. 
\section{Generalizations}\label{gensec}
\subsection{Higher dimensions}\label{higherdim}
We can easily generalize these results to higher dimensions. We start with higher dimensional versions of the Euclidean solutions (\ref{euclschw}) and (\ref{euclhyper}):

\be
ds^2 = f(r)d\chi^2 + \f{dr^2}{f(r)}+r^2 \left(d\h^2+\sin^2 \h d\Omega_{d-3}^2+\cos^2 \h d\phi^2\right)\,, \qquad f(r)={r^2} +1-\f{m}{r^{d-2}}.\label{spherehigh}
\ee
and
\be
ds^2 = g(\rho)d\phi^2 + \f{d\rho^2}{g(\rho)}+\rho^2\, \f{d\chi^2 + d \tilde\theta^2+\sin^2\tilde \h d\Omega_{d-3}^2}{\cos^2 \tilde \h}\,, \qquad g(\rho)={\rho^2} -1-\f{\mu}{\rho^{d-2}}.\label{hyperhigh}
\ee
The restrictions on the parameters $m, \mu$ and radial coordinates are similar to before. The coordinate $\h$ now lies in  $0\le \h\le \pi/2$. The natural metric on the conformal boundary is $S^1 \times S^{d-1}$ in the first case and $S^1\times\mathcal{H}^{d-1}/\mathbb{Z}$ in the second (assuming we compactify $\chi$ as before). However, by applying the coordinate transformation (\ref{hypercoord}) to \eqref{spherehigh} we obtain a natural boundary metric $S^1\times\mathcal{H}^{d-1}/\mathbb{Z}$, and applying the coordinate transformation (\ref{sphercoord}) to \eqref{hyperhigh} we obtain a natural boundary metric $S^1 \times S^{d-1}$ (possibly with conical singularities). 

In even dimensions, we will be faced with the new feature of a conformal anomaly due to our nontrivial Weyl transformation. This anomaly will enter when transforming individual free energies or correlation functions from one frame to another. Since the contribution of the anomaly depends only on the Weyl transformation, the confining phase transition on $S^1 \times S^{d-1}$ will determine the location of the confining phase transition on $S^1 \times \hz$ without an additional anomaly contribution.

\subsection{Detaching circles from other manifolds}
Here we will illustrate that the notion of detachable circles can be used to provide a new viewpoint for finite temperature theories on other manifolds. A circle is detachable whenever there is a hypersurface orthogonal Killing vector with compact orbits. Here we will focus on the cases $\mathcal{H}^{d-1}$ and $\mathbb{R}^{d-1}$ due to their physical relevance. The $S^1 \times \mathcal{H}^{d-1}$ case is physically relevant due to the fact that $\log Z[S^1 \times \mathcal{H}^{d-1}]=\log Z[S^3]$ in a conformal theory computes the entanglement across a disc in flat space, which is monotonic under renormalization group flow for $d\leq 4$ and proposed to be monotonic for $d>4$. The $S^1 \times \mathbb{R}^{d-1}$ case is physically relevant since it captures high-temperature physics of local theories on generic manifolds.

Consider detaching circles in the two cases as follows:
\begin{align}\label{hypercircle}
\hspace{-4mm}S^1 \times \mathcal{H}^{d-1}\hspace{-1mm}&: d\chi^2 + d\rho^2 + \sinh^2\hspace{-1mm}\rho \,d\phi^2 + \cosh^2\hspace{-1mm}\rho\, d\mathcal{H}_{d-3}^2&&\hspace{-3mm}\longrightarrow \, \f{d\chi^2+d\rho^2 + \cosh^2 \hspace{-1mm}\rho\, d\mathcal{H}_{d-3}^2}{\sinh^2\hspace{-1mm} \rho}+d\phi^2\\
\label{flatcircle}\hspace{-40mm}S^1 \times \mathbb{R}^{d-1}\hspace{-1mm}&: d\chi^2 + dr^2 + \hspace{7mm}dx_i^2 \hspace{9.2mm}+ \hspace{7.2mm}r^2\, d\phi^2 &&\hspace{-3mm}\longrightarrow \,\f{d\chi^2+dr^2 + dx_i^2}{r^2}+d\phi^2\,,
\end{align}
where $d\mathcal{H}_{d-3}$ is a global metric on $(d-3)$-dimensional hyperbolic space. The right-hand-side of \eqref{hypercircle} is again $S^1 \times \mathcal{H}^{d-1}$, which can be seen by defining $\sinh y =\sinh^{-1}\rho$, after which the metric becomes  
\be
ds^2=\sinh^2 y \,d\chi^2+ dy^2 + \cosh^2 y \,d\mathcal{H}_{d-3}^2+d\phi^2\,.
\ee 
To have a smooth hyperboloid, we need $\chi \sim \chi + 2\pi$. More generally, \eqref{hypercircle} shows that a CFT on a conical hyperbolic space with 
$\chi \sim \chi + L_\chi$ and at inverse temperature $2\pi$ is equivalent to the CFT on the smooth hyperboloid at  inverse temperature $L_\chi$.  If $L_\chi = 2\pi$, the conical singularity disappears and the spacetime is invariant under this Weyl transformation. Notice that this transformation is not trivial: it will constrain correlation functions since local operators will transform nontrivially under such a transformation, but this should simply repackage known constraints from conformal invariance on the plane.

The right-hand-side of \eqref{flatcircle} is $S^1 \times \mathcal{H}^{d-1}/\mathbb{Z}$, but the identification is qualitatively different than the one we have considered so far. 
There are actually  three inequivalent ways to quotient $\mathcal{H}^{d-1}$. This can be seen by embedding $\mathcal{H}^{d-1}$ in $d$-dimensional Minkowski space. One can clearly identify points under a boost, a rotation, or a null boost, i.e. a boost generated by $F = \ell \wedge v$, where $\ell$ is null. The identification we have been considering so far corresponds to a boost. The result is a noncontractible circle with a minimum length.

In the $S^1 \times \mathbb{R}^{d-1}$ case, we obtain a quotient of hyperbolic space by a null boost. There is now a noncontractible circle which becomes arbitrarily small. Changing the periodicity of $\chi$ does not change this geometry since it can be absorbed by a rescaling of $r$ and $x_i$. As before, we obtain the CFT on this hyperbolic quotient at inverse temperature $2\pi$. The theory on this hyperbolic space at arbitrary temperature $\phi \sim \phi + 2\pi \alpha$ provides a definition of the orbifold flat space theory. This transformation between a locally flat space and a space that is locally $S^1 \times \mathcal{H}^{d-1}$ is different than the Casini-Huerta-Myers map used to calculate entanglement entropies using the replica trick \cite{Casini:2011kv}. 

 We showed in section 3 that a CFT on  $\hy^{d-1}/\mathbb{Z}$ (the quotient under a boost) has a phase transition. 
We now claim that there is a similar thermal phase transition on the space $(d\chi^2+dr^2 + dx_i^2)/r^2$ (the quotient under a null boost). Consider a CFT on this space at temperature $T= 1/2\pi$. The natural bulk dual is the $\mu = 0$ hyperbolic black hole with this identification of the hyperboloids. There is another  bulk solution which can be obtained by starting with the planar black hole
\be
ds^2 = \f{1}{z^2}\left [( 1 - m z^3 )d\chi^2 + ( 1- m z^3)^{-1} {dz^2}+d r^2 + r^2 d\phi^2\right]
\ee
and introducing a new radial coordinate $\rho = r/z$. The natural boundary metric becomes $(d\chi^2 + dr^2)/r^2 + d\phi^2$. To see which solution dominates the canonical ensemble, we need to compare the actions of these two solutions. But since the first is locally AdS, it is equivalent to the $m=0$ planar black hole. It is now clear that the $m>0$ solution will dominate. Since the $\chi$ circle caps off in the bulk, this is like a confining phase. If we now increase the temperature, we expect to enter a deconfined phase where the bulk dual is described by the usual hyperbolic black hole with $\mu > 0$. 

\subsection{Adding angular momentum}
We now relax the requirement that the detachable circle must be hypersurface orthogonal, and ask what happens to our transformation if we add rotation. There is a rotating generalization of the Hawking-Page transition which should describe a rotating extension of the phase transition we have found on $\mathcal{H}^{2}/\mathbb{Z}$.

The Kerr-AdS metric  (with the AdS radius again set to unity) is \cite{Carter:1968ks}
\be
ds^2 = -\f{\Delta_r}{\Sigma^2} \left [ dt - \f{a\sin^2\h}{1-a^2} d\phi\right]^2 + \f{\Sigma^2}{\Delta_r} dr^2 + \f{\Sigma^2}{\Delta_\theta} d\h^2 + \f  {\Delta_\theta \sin^2\h}{\Sigma^2}
\left[adt - \f{r^2 + a^2}{1-a^2} d\phi\right]^2
\ee
where
\be
\Delta_r = (r^2+a^2)(1+r^2) - 2mr,\quad \Delta_\theta = 1-a^2\cos^2 \theta, \quad \Sigma^2 = r^2 + a^2\cos^2 \theta
\ee
The parameter $a$ controls the angular momentum and is restricted to $a^2 < 1$. The horizon is located at $r=r_+$, the largest root of $\Delta_r(r)$. The inverse temperature is
\be
\beta = \f{4\pi r_+ (r_+^2 + a^2)}{3r_+^4 +(1+a^2)r_+^2 -a^2}
\ee
To obtain the standard boundary metric, it is convenient to define new coordinates  \cite{Hawking:1998kw}
\be
T = t,\quad \Phi = \phi + at,\quad y\cos\Theta = r\cos\theta,\quad 
y^2 = \f{r^2 \Delta_\theta  +a^2\sin^2\theta}{1-a^2}
\ee
The conformal boundary then takes the usual form
\be 
ds^2 = -dT^2 + d\Theta^2 +\sin^2\Theta d\Phi^2
\ee
and the angular velocity of the black hole relative to this static frame at infinity is
\be
\Omega = a \f{r^2_+ +1}{r^2_+ +a^2}
\ee 
To obtain a real Euclidean solution, we analytically continue $t = -i\chi$ and $a=i\alpha$. Setting 
\be
\tilde \Omega = \f {\alpha (1+\alpha^2)}{r_+^2 -\alpha^2}
\ee
one finds that the Killing vector $\xi = \p_\chi + \tilde \Omega \p_\phi$ vanishes at $r=r_+$,  and regularity at this point requires that we identify $(\chi,\phi)$ with $(\chi +L_\chi,\phi + L_\chi \tilde \Omega)$, where $L_\chi = \beta$ to conform with past notation. Setting $\tilde \phi = \phi - \tilde\Omega \chi$, we see that $\tilde \phi $ is constant along the orbits of $\xi$. So in terms of $(\chi, \tilde\phi)$ coordinates, the identification is simply $(\chi, \tilde\phi) = (\chi + L_\chi, \tilde \phi )=(\chi,\tilde\phi+2\pi)$. In terms of the boundary geometry this becomes
\be\label{bdyrot}
ds^2 = d\chi^2 + d\Theta^2 +\sin^2\Theta [d\tilde \phi + (\tilde\Omega+ \alpha) d\chi]^2\,,
\ee
where $\tilde{\Omega}+\alpha = -i \Omega$. The Euclidean action, relative to pure AdS, changes sign when $r_+ = 1$ \cite{Hawking:1998kw}. This implies there is a phase transition directly analogous to the original Hawking-Page transition.

By choosing appropriate hyperbolic slices of this solution we can rescale \eqref{bdyrot} by $\sin^{-2}\Theta$. The result is a twisted product of $S^1$ and $\mathcal{H}^{2}/\mathbb{Z}$. In analogy with our earlier discussion, we can conclude that this space has a phase transition at inverse temperature $2\pi$ when one varies the size of the circle $L_\chi$.      
The holographic stress tensor of this transformed solution can be calculated as in section \ref{somecalc}, with the result that the energy density and angular momentum density (with respect to $\tilde{\phi}$ as time) vanish as $\sin^3\Theta$ as $\Theta \rightarrow 0$ and $\Theta \rightarrow \pi$. This gives a finite total energy and angular momentum. The solution generalizes our transformed Schwarzschild-AdS solution by adding localized angular momentum to the localized energy deformation on the hyperboloid. The fact that the angular velocity is not constant in the transformed solution is related to the fact that the $[d\tilde{\phi}+(\tilde{\Omega}+\alpha)d\chi]^2$ term does not appear inside the $\mathcal{H}^2/\mathbb{Z}$, i.e. does not appear with a factor of $\sin^{-2}\Theta$.

\subsection{Returning to $S^{d-1}$}\label{startriangle}
Throughout this paper we have discussed a duality which relates CFT$_d$ on $S^{d-1}$ at a given temperature to the theory on $\hz$ at some inversely related temperature. It is natural to wonder if the theory on $\hz$ can then be related back to the theory on $S^{d-1}$ by some other method that is not just inverting the original duality, since that would be contentless. As discussed in the introduction, this is conceptually analogous to the star-triangle relation used to relate the Ising model on the honeycomb lattice to the the Ising model on the triangle lattice. When composed with Kramers-Wannier duality, this gives a self-duality relation on the triangular lattice. 

In our case, we can make some progress by using the universality of high-temperature physics. For example, if asymptotically low temperature on $S^{d-1}$ was related to asymptotically high temperature on $\hz$, then we could use the fact that at leading order at asymptotically high temperature the curvature of the space does not enter into e.g. the free energy. However, our temperature-inversion relation is not so simple: for a smooth $S^{d-1}$ the inverse temperature on $\hz$ is fixed to be $2\pi$ in units of the radius of curvature of the hyperboloid. One can potentially make progress by considering the orbifold $S^{d-1}_\alpha$ and shrinking the azimuthal angle, so that in the hyperbolic frame we are at asymptotically high temperature. Since from our point of view the orbifold theory is \emph{defined} by the hyperbolic frame, any resulting self-duality relation on $S^{d-1}_\alpha$ would have to be understood as a self-duality relation on $\hz$. We will not consider this approach.

We can instead use an effective theory argument introduced in \cite{meup}, which requires restricting to theories which are gapped at finite temperature. For such a theory, we can dimensionally reduce over the thermal circle and write an effective theory on the spatial manifold. By gappedness, the resulting theory is a local theory in the available background terms, which in CFT$_d$ are simply curvature terms. The expansion is in the Wilsonian scale $\beta$, the inverse temperature. So we have
\be
\log Z(\beta) = \int d^{d-1}x\, \sqrt{h}\left(c_1\f{1}{\beta^{d-1}}+c_2\f{R^{(1)}}{\beta^{d-3}}+c_3 \f{R^{(2)}}{\beta^{d-5}}+\dots\right)
\ee
This is to be understood as a perturbative expansion around $\beta \rightarrow 0$. Powers $R^{(n)}$ are to be understood as all possible combinations of contractions of the Riemann tensor 
for the $(d-1)$-dimensional metric $h_{\mu\nu}$, with e.g. different coefficients between $R_{\mu\nu}R^{\mu\nu}$ and $R^2$ which are suppressed for simplicity. For an effective theory capturing correlation functions we also have the background sources available, but for now we restrict to this case. 

As discussed in \cite{meup}, this effective theory makes clear that the high-temperature 
free energy density on a hyperboloid is related to the high-temperature free energy density on the sphere simply by sign flips in the terms of the effective theory with odd powers of curvature. This is easily checked to be true for example in holographic theories. Since this effective theory is valid as a perturbative expansion around $\beta \rightarrow 0$, we cannot justifiably use it in our context even for CFTs gapped at finite temperature, since the theory on the hyperboloid is at inverse temperature $2\pi$. However, for certain large-$N$ theories (a concrete example being holographic theories, to which we now restrict), the nonperturbative corrections which would spoil such an effective theory beyond $\beta \rightarrow 0$ are suppressed.\footnote{Holographic theories of course have corrections in $Z$ which are suppressed exponentially in $N$; these are just the existence of distinct bulk saddles. Here we are referring to corrections $e^{-1/\beta}$ in $\log Z$ when expanded about high temperature (see e.g. \cite{Kutasov:2000td}), which generically exist in free theories but are absent in strongly coupled holographic theories.} This means that this effective theory has an ``extended range of validity" up to some $O(1)$ temperature which for holographic theories corresponds to the Hawking-Page phase transition. Above this temperature we are in a deconfined phase for which the effective theory is valid. 

To be concrete, a low-temperature confined phase on $S^{d-1}$ maps to a high-temperature $(\beta = 2\pi)$ deconfined phase on $\hz$. But as just argued this high-temperature deconfined phase is captured by an effective theory on $\hz$, which is simply related to the effective theory on $S^{d-1}$. In this way, we have a high-temperature/low-temperature duality for the theory on $S^{d-1}$! 
A problem with implementing this for holographic theories is that the theory with $\beta = 2\pi$ on $S^{d-1}$ with $d>2$ is always in the confined phase, i.e. the effective field theory written above is invalid at such low temperatures. One could circumvent this problem by understanding the low-temperature confined phase on $S^{d-1}_\alpha$ perturbatively in $2\pi(1 - \alpha)$. This would map to $\beta = 2\pi \alpha$ on $\hz$. Using analyticity, one could then extract the free energy density at arbitrary $\beta < 2\pi$. The effective theory could then relate this free energy density to that on $S^{d-1}$, where we can now safely pick a sufficiently small $\beta$ such that the effective theory on $S^{d-1}$ is valid.

\section{Open questions}

There are several open questions raised by this work which remain to be investigated. 
One is to find smooth bulk solutions to complete the phase diagram in figure \ref{phasediag}. The basic problem is that the hyperbolic solutions (\ref{euclhyper}) or \eqref{hyperhigh} contain two adjustable parameters ($L_\phi$ and $L_\chi$) while the spherical solutions contain only one, $L_\chi$. A smooth generalization of the hyperbolic black holes has recently been found \cite{Chen:2015zoa} in four dimensions by using the AdS C-metric. It would be interesting to explore whether they can be used to describe new phases in  the phase diagram.

A related question is to find smooth bulk solutions to describe other phases of a CFT on the twisted product space discussed in section 6.3. Since we started with the Kerr-AdS metric, one would like an analogous solution based on the hyperbolic black hole.  There is indeed a ``rotating hyperbolic black hole" \cite{Klemm:1997ea} constructed by analytically continuing 
\be
t \to it,\quad r\to i\rho, \quad \theta \to i\psi, \quad m\to i \mu, \quad a = i\alpha
\ee
in the Kerr-AdS metric. But this metric does not appear to have a boundary geometry conformal to \eqref{bdyrot}.

Another interesting question to explore is whether there is an emergent $SL(2,\mathbb{Z})$ invariance in the limit of a large-order quotient, i.e. whether there is a modular equivalence between $S^1_{2\pi/p} \times S^{d-1}_\alpha$ and $S^1_{2\pi \alpha} \times S^{d-1}_{p^{-1}}$ in the limit of large $p$ and small $\alpha$. The intuitive justification for this emergent invariance in the case of smooth quotients \cite{Shaghoulian:2016gol} -- Gromov collapse and properties of fundamental groups --  exists for orbifolds and operates in a similar way as in the case of manifolds \cite{thurstonbook, orbifoldbook, Dunbar1988}. The relation we have exhibited in this paper between $S^1\times S^{d-1}_\alpha$ and $S^1 \times \hy^{d-1}/\mathbb{Z}$ may serve as a starting point for arguing for such an equivalence.  

Yet another open question is whether the divergence we have seen (in section 4.1) in the energy density near the conical singularities on $S^2_\alpha$ is universal. If so, one could perhaps define a finite energy for states on the orbifold  by subtracting a universal divergent term. 

Finally, while we have outlined a temperature-inversion duality on $S^{d-1}$ in section \ref{startriangle}, it remains to use this to calculate finite-temperature observables in theories satisfying the assumptions in that section. 


\vskip 1cm
\centerline{\bf Acknowledgements}
\vskip .5cm 
It is a pleasure to thank Thomas Hartman and Don Marolf for discussions. This work was supported in part by NSF grant no. PHY-1504541 and PHY-1316748, and Simons Foundation Grant 488643. The work of ES was performed in part at the Aspen Center for Physics, which is supported by NSF grant no. PHY-1607611.

\footnotesize
\bibliographystyle{apsrev4-1long}
\bibliography{orbifoldsBIB}

\end{document}